\documentstyle [12pt,a4,epsf] {article}

\newcommand{\e} { {\rm e} }
\newcommand{\eg} {{\it e.g.}}
\newcommand{\eq} { {\rm eq} }
\newcommand{\ie} {{\it i.e.}}
\renewcommand{\pd} {\partial}

\begin{document}


\title {Adsorption Kinetics of Surfactants \\
            at Fluid-Fluid Interfaces}
\author {Haim Diamant and David Andelman \\
         School of Physics and Astronomy   \\
         Raymond and Beverly Sackler Faculty of Exact Sciences \\
         Tel Aviv University, Ramat Aviv, 69978 Tel Aviv, Israel \\
         \\}
\date{October 1996}

\maketitle


\begin{abstract}
\setlength {\baselineskip} {10pt}

We review a new theoretical approach to the kinetics of 
surfactant adsorption at fluid-fluid interfaces.
It yields a more complete description of the kinetics both
in the aqueous solution and at the interface, deriving
all equations from a free-energy functional.
It also provides a general method to calculate dynamic
surface tensions.
For non-ionic surfactants the results coincide with 
previous models. Common non-ionic surfactants are shown
to undergo diffusion-limited adsorption, in agreement 
with experiments.
Strong electrostatic interactions in salt-free ionic 
surfactant solutions are found to lead to kinetically 
limited adsorption.
In this case the theory accounts for unusual experimental
results which could not be understood using previous 
approaches.
Added salt screens the electrostatic interactions and 
makes the ionic surfactant adsorption similar to the non-ionic
case.
The departure from the non-ionic behavior as the salt 
concentration is decreased is calculated perturbatively.

\end{abstract}


\pagebreak
\section{Introduction}
\label{introduction}
\setcounter{equation}{0}
\setlength {\baselineskip} {10pt}

The kinetics of surfactant adsorption is a fundamental problem 
of interfacial science playing a key role in various processes and
phenomena, such as wetting, foaming and stabilization of liquid
films.
Since the pioneering theoretical work of Ward and Tordai in the 
1940s \cite{WT}, it has been the object of thorough 
experimental and theoretical research \cite{review}.

The problem being a non-equilibrium one, a few theoretical 
questions immediately arise. 
One question concerns the kinetic adsorption mechanism to be 
employed by the model.
One might assume a sort of an equilibrium adsorption isotherm to 
hold at the interface (\eg, as in \cite{Delahay}--\cite{Joos1}), 
or, alternatively, use a full kinetic equation 
(\eg, \cite{Miller}--\cite{Franses}).
Another important question relates to the definition and 
calculation of the time-dependent interfacial tension as
measured in experiments.

Previous theoretical works have addressed these questions 
by adding appropriate assumptions to the theory. 
Such models can be roughly summarized by the following 
scheme:
(i) consider a diffusive transport of surfactant molecules from
a semi-infinite bulk solution (following Ward and Tordai);
(ii) introduce a certain adsorption equation as a boundary condition
at the interface;
(iii) solve for the time-dependent surface coverage;
(iv) assume that the equilibrium equation of state is valid
also out of equilibrium and calculate the dynamic surface tension
\cite{Fordham}.

In the current paper we would like to review an alternative
approach based on a free-energy formalism \cite{EPL,JPC}.
The main advantage is that all the equations 
are derived from a single functional, thus yielding a more complete
and consistent description of the kinetics in the entire system.
Results of previous models can be recovered as special cases, and
one can check the conditions under which such cases hold.
The definition and calculation of the dynamic surface tension 
results naturally from the formalism itself, and extension to
more complicated interactions can follow.

We restrict ourselves in the current paper to a simple, yet 
rather general case.
A sharp, flat interface is assumed to separate an aqueous 
surfactant solution from another fluid, non-polar phase.
The solution is assumed to be below the critical micelle
concentration, \ie, it contains only monomers.
We start in Section~\ref{nonionic} by considering the adsorption of 
non-ionic surfactants, for which previous theories yield satisfactory
results.
We then proceed in Section~\ref{ionicnosalt} to discuss salt-free
ionic surfactant solutions, where strong electrostatic interactions exist
and interesting time dependence has been observed in experiments
\cite{Langevin}.
In Section~\ref{ionicsalt} the effect of added salt to ionic surfactant
solutions is examined.

We shall not describe various experimental techniques which have
been devised in the context of adsorption kinetics of surfactants.
Such information can be found in Ref.~\cite{review} and in the 
contribution by A. Pitt included in this volume.

\section{Non-Ionic Surfactants}
\label{nonionic}

We identify the measurable change in interfacial tension, 
$\Delta\gamma$, with the excess in free energy per unit area due 
to the adsorption at the interface.
This definition is assumed to hold both at equilibrium and out of 
equilibrium.
The free energy excess can be written as a functional of the volume 
fraction profile of the surfactant, $\phi(x,t)$,  $x$ being the distance 
from the interface and $t$ the time,
\begin{equation}
  \Delta \gamma [\phi] = \int_0^\infty \Delta f[\phi(x,t)] dx,
 \label{Dg}
\end{equation}
where $\Delta f$ is the local excess in free energy density
over the bulk, uniform solution.

We take the bulk solution to be dilute and assume a contact with a 
reservoir, where the surfactant has fixed volume fraction
and chemical potential, $\phi_b$ and $\mu_b$, respectively.
Steric and other short-range interactions between surfactant
molecules are assumed to take place only within a molecular distance 
from the interface.
This is motivated by the observation that the profile of a soluble surfactant
monolayer is in practice almost ``step-like'', the volume fraction at the
interface itself being many orders of magnitudes larger than that 
in the solution.

Hence, we write the local free energy density as
\begin{eqnarray}
  \Delta f &=& (T/a^3) [ \phi(\ln\phi - 1) - 
                     \phi_b(\ln\phi_b - 1) ] - \mu_b (\phi - \phi_b)
  \nonumber \\
               & & + \{ (T/a^2) [ \phi\ln\phi + (1-\phi)\ln(1-\phi) ] - 
                      (\alpha + \mu_1)\phi - (\beta/2)\phi^2 \} a \delta(x),
 \label{Df}
\end{eqnarray}
where $a$ denotes the surfactant molecular dimension and $T$ the 
temperature (taking the Boltzmann constant as 1).
Note that this functional divides the system into 
two distinct, coupled sub-systems --- the bulk solution and the interface
\cite{Tsonop}.
As a result we shall obtain distinct equilibrium and kinetic equations
for these two sub-systems.
The contribution from the bulk contains only the ideal entropy of mixing 
in the dilute solution limit and contact with the reservoir.
In the interfacial contribution we have included the entropy
of mixing accounting for the finite molecular size, a linear term
accounting for the surface activity and contact with the adjacent
solution [$\mu_1\equiv\mu(x\rightarrow 0)$ being the chemical 
potential at the adjacent layer],
and a quadratic term describing short-range lateral attraction 
between surfactant molecules at the interface.
The surface activity parameter, $\alpha$,  is typically of order $10T$,
and the lateral attraction parameter, $\beta$, is typically a few $T$.

Although the functional  (\ref{Df}) has a simple form, it yields
physically non-trivial results. 
More complicated cases, \eg, certain surfactants whose adsorption
seem to be hindered by a potential barrier, may  require additional terms.
Such terms, however, can be easily incorporated, as demonstrated in the 
next section for electrostatic interactions.

{\em {\bf Equilibrium relations}} are readily obtained by setting the variation
of the free energy with respect to $\phi(x)$ to zero,
\[
  \frac {\delta\Delta\gamma} {\delta\phi(x)} = 0, \ \ \ \ \ \mbox{equilibrium}.
\]
This yields in the current simple case a uniform profile in the bulk,
$\phi(x>0) \equiv \phi_b$, and recovers the Frumkin adsorption
isotherm (or the Langmuir one, if $\beta=0$) \cite{Adamson} 
at the interface,
\begin{equation}
  \phi_0 = \frac {\phi_b} {\phi_b + \e^{-(\alpha+\beta\phi_0)/T}},
 \label{Frumkin}
\end{equation}
where $\phi_0\equiv\phi(x=0)$ denotes the surface coverage.
Substituting these results in the free energy functional recovers also
the equilibrium equation of state,
\begin{equation}
  \Delta\gamma = [ T\ln(1-\phi_0) + (\beta/2) \phi_0^2 ] / a^2.
 \label{eqstate}
\end{equation}

{\em\bf {Kinetic equations}} can also be derived from the variation of 
the free energy.
The conventional scheme in the case of a conserved order parameter is
\cite{Langer}
\[
   \frac{\pd\phi}{\pd t} = (a^3 D/T) \frac{\pd}{\pd x} \left[
   \phi \frac{\pd}{\pd x} \left( \frac{\delta\Delta\gamma}{\delta\phi}
   \right) \right],
\]
where $D$ is the surfactant diffusion coefficient.
This leads to an ordinary diffusion equation in the bulk,
\begin{equation}
  \frac{\pd\phi}{\pd t} = D \frac{\pd^2\phi}{\pd x^2},
 \label{diffusion}
\end{equation}
and to a conservation condition at the layer adjacent to the interface,
\begin{equation}
  \frac{\pd\phi_1}{\pd t} = (D/a) \left.
        \frac{\pd\phi}{\pd x} \right|_{x=a}
        - \frac{\pd\phi_0}{\pd t},
 \label{dp1dt}
\end{equation}
where $\phi_1\equiv\phi(x\rightarrow 0)$ is the local volume fraction.
Finally, at the interface itself, we get
\begin{equation}
  \frac{\pd\phi_0}{\pd t} = (D/a^2) \phi_1
          \left[ \ln \frac{\phi_1(1-\phi_0)}{\phi_0} + \frac{\alpha}{T}
          + \frac{\beta\phi_0}{T} \right].
 \label{dp0dt}
\end{equation}
We have assumed, for simplicity, that the surfactant diffusion coefficient,
$D$, is the same in the bulk and near the interface in spite of the different
environments.
In reality this should not be strictly accurate.

Our formalism has led to a diffusive transport
in the bulk [Eqs.~(\ref{diffusion}) and (\ref{dp1dt})] coupled to
an adsorption mechanism at the interface [Eq.~(\ref{dp0dt})].
Yet unlike previous models, all of the equations have been derived
from a single functional, and hence, various assumptions employed
by previous works can be examined.
Treating Eqs.~(\ref{diffusion}) and (\ref{dp1dt}) using
the Laplace transform with respect to time, we obtain a relation
similar to the Ward and Tordai result \cite{WT},
\begin{equation}
  \phi_0(t) = (\sqrt{D/\pi}/a) \left[
          2\phi_b\sqrt{t} - \int_0^t \frac {\phi_1(\tau)} 
          {\sqrt{t-\tau}} d\tau \right]
          + 2\phi_b - \phi_1,
 \label{WT}
\end{equation}
with a small difference coming from the finite thickness we have 
assigned to the sub-surface layer of solution 
(vanishing for $a\rightarrow 0$).

The diffusive transport from the bulk solution [Eq.~(\ref{WT})]
relaxes like
\begin{eqnarray}
  \phi_1(t)/\phi_b &\simeq& 1 - \sqrt{\tau_d/t}; \ \ \ \ \ t\rightarrow\infty
    \nonumber \\
  \tau_d &\equiv& (a^2/\pi D)(\phi_{0,{\eq}}/\phi_b)^2,
 \label{asymdiff}
\end{eqnarray}
where $\phi_{0,{\eq}}$ denotes the equilibrium surface coverage.
The molecular diffusion time scale, $a^2/D$, is of order $10^{-9}$~sec,
but the factor $\phi_{0,{\eq}}/\phi_b$ in surfactant monolayers
is very large (typically $10^5-10^6$), so the diffusive transport to
the interface may require minutes.
The kinetic process at the interface [Eq.~(\ref{dp0dt})] relaxes like
\begin{eqnarray}
  \phi_0(t)/\phi_{0,{\eq}} &\simeq& 1 - \e^{-t/\tau_k} \ \ \ \ \ 
      t \rightarrow \infty    \nonumber \\
  \tau_k &\equiv& (a^2/D)(\phi_{0,{\eq}}/\phi_b)^2 
     \e^{-(\alpha+\beta\phi_{0,{\eq}})/T}.
 \label{asymkin}
\end{eqnarray}
Since $\alpha$ for common surfactants is of order $10T$, we expect $\tau_k$
to be much smaller than $\tau_d$.
In other words, the adsorption of common non-ionic surfactants, not
hindered by any high potential barrier, is expected to be 
{\em diffusion-limited}.
The asymptotic time dependence (\ref{asymdiff}) yields a distinct
``footprint'' for diffusion-limited adsorption, as demonstrated in
Fig.~1.

In mathematical terms the adsorption being diffusion-limited means
that the variation of the free energy with respect to $\phi_0$
can be neglected at all times whereas the variation with respect to $\phi(x>0)$
cannot.
This has two consequences.
The first is that the relation between $\phi_0$ and $\phi_1$ is given at
all times by the equilibrium adsorption isotherm [(\ref{Frumkin}) in
our model].
The solution of the adsorption problem in the non-ionic,
diffusion-limited case amounts, therefore, to the 
simultaneous solution of the Ward-Tordai equation (\ref{WT})
and the adsorption isotherm.
Exact analytical solution exists only for the simplest, linear
isotherm, $\phi_0\propto\phi_1$ \cite{Sutherland}.
For more realistic isotherms such as (\ref{Frumkin}), one has
to resort to numerical techniques (useful numerical schemes can 
be found in Refs.~\cite{review,Lin1}).
The second consequence of the vanishing of 
$\delta\Delta\gamma/\delta\phi_0$  is that the dynamic surface tension, 
$\Delta\gamma(t)$, approximately obeys the equilibrium equation of 
state (\ref{eqstate}).
These two consequences show that the validity of the schemes 
employed by previous theories is essentially restricted to 
diffusion-limited cases.

The dependence defined by the equilibrium equation of state
(\ref{eqstate}) is depicted in Fig.~2a.
As a result of the competition between the entropy and interaction 
terms in Eq.~(\ref{eqstate}) the surface tension changes very little
for small surface coverages.
As the coverage increases beyond about $1-(\beta/T)^{-1/2}$, the 
surface tension starts decreasing until reaching equilibrium.
This qualitatively explains the shape of dynamic surface tension 
curves found in experiments for non-ionic surfactants (\eg,
\cite{Lin1,Lin2}).
We have reproduced in Fig.~2b one such curve published by Lin 
{\it et al.} \cite{Lin1}.
The theoretical solid curve was obtained by these writers using a
scheme similar to the one just described ---
solution of the Ward-Tordai equation together with the Frumkin
isotherm and substitution in the equation of state to calculate the
surface tension.
Note that the parameters $\alpha$, $\beta$ and $a$ can be fitted
from independent equilibrium measurements, so the dynamic
surface tension curve has only one fitting parameter, namely the
diffusion coefficient, $D$.
As can be seen, the agreement with experiment is quite satisfactory.
However, when the adsorption is not diffusion-limited, such a 
theoretical approach is no longer applicable, as will be demonstrated 
in the next section.

\section{Salt-Free Ionic Surfactant Solutions}
\label{ionicnosalt}

We turn to the more complicated but important problem of ionic surfactant
adsorption, and start with the salt-free case where strong 
electrostatic interactions are present.
In Fig.~3 we have reproduced experimental results published by
Bonfillon-Colin {\it et al.} for SDS solutions with (open circles) 
and without (full circles) added salt \cite{Langevin}.
The salt-free ionic case exhibits a much longer process with
a peculiar intermediate plateau.
Similar results were presented by Hua and Rosen for DESS 
solutions \cite{HuaDESS}.
A few theoretical models were suggested for the problem of
ionic surfactant adsorption \cite{Dukhin,Borwasan2,Radke},
yet none of them could produce such dynamic surface
tension curves.
It is also rather clear that a theoretical scheme such as the one discussed 
in the previous section cannot fit these experimental results.
On the other hand, addition of salt to the solution screens the electrostatic 
interactions and leads to a behavior very similar to the non-ionic one.
We shall return to this issue in Sec.~\ref{ionicsalt}.
We thus infer that strong electrostatic interactions 
affect drastically the adsorption kinetics.
Let us now study this effect in more detail. 
We follow the same lines presented in the previous section while
adding appropriate terms to account for the additional interactions.

Our free energy functional in the salt-free ionic
case is divided into three contributions: a contribution from the 
surfactant, one from the counterions and one from the electrostatic 
field.
It depends on three degrees of freedom:
the surfactant profile, $\phi^+(x,t)$ (we take the surfactant
ion to be the positive one), the counterion
profile, $\phi^-(x,t)$, and a mean electric potential,
$\psi(x,t)$.
\begin{equation}
  \Delta \gamma [\phi^+,\phi^-,\psi] = \int_0^\infty 
        [ \Delta f^+(\phi^+) + \Delta f^-(\phi^-) + f_{\rm el}(\phi^+,\phi^-,\psi) ] dx.
 \label{Dg2}
\end{equation}
The surfactant contribution, $\Delta f^+$, is identical to Eq.~(\ref{Df})
of the non-ionic case.
In the counterion contribution, $\Delta f^-$, we include only the
bulk part of Eq.~(\ref{Df}), taking the counterions at this stage to be
completely surface-inactive.
The electrostatic contribution contains interactions between the ions and
the electric field and the energy stored in the field itself,
\begin{equation}
  f_{\rm el} = e \left( \frac{\phi^+}{(a^+)^3} - \frac{\phi^-}{(a^-)^3} 
           \right) \psi - \frac{\varepsilon}{8\pi} \left( \frac{\pd\psi}
           {\pd x} \right)^2 + \frac{e}{(a^+)^2}\phi^+\psi\delta(x),
 \label{fel}
\end{equation}
where $a^\pm$ are the molecular sizes of the two ions, $e$ the 
electronic charge and $\varepsilon$ the dielectric constant of water.
For simplicity we have restricted ourselves to fully ionized, monovalent
ions [which implies that $\phi^+_b/(a^+)^3=\phi^-_b/(a^-)^3= c_b$,
$c_b$ being the bulk concentration].

Ions in solution, apart from interacting with other ions, also feel 
repulsion from the interface due to ``image-charge'' effects, as
discussed by Onsager and Samaras \cite{Onsager}.
It can be shown, however, that these effects become negligible
as soon as the surface coverage exceeds about 2 percents
\cite{JPC}.

{\em \bf {Equilibrium equations}} are readily obtained, as in the previous section,
by setting the variation of the free energy with respect to the various degrees
of freedom to zero,
\[
  \frac {\delta\Delta\gamma} {\delta\phi^\pm(x)} = 
  \frac {\delta\Delta\gamma} {\delta\psi(x)} = 0, \ \ \ \ \ \mbox{equilibrium}.
\]
These equations yield the Boltzmann ion profiles,
\[ 
  \phi^\pm(x>0) = \phi^\pm_b \e^{\mp e\psi(x)/T},
\]
the Poisson equation,
\begin{equation}
  \frac{\pd^2\psi}{\pd x^2} = -\frac{4\pi e}{\varepsilon}
  \left( \frac{\phi^+}{(a^+)^3} - \frac{\phi^-}{(a^-)^3} \right),
 \label{Poisson}
\end{equation}
the electrostatic boundary condition,
\begin{equation}
  \left. \frac{\pd\psi}{\pd x} \right|_{x=0} =
       -\frac{4\pi e}{\varepsilon (a^+)^2} \phi^+_0,
 \label{neutral}
\end{equation}
and, finally, recovers the Davies adsorption isotherm \cite{Davies},
\begin{equation}
  \phi^+_0 = \frac {\phi^+_b} {\phi^+_b + 
           \e^{-(\alpha+\beta\phi^+_0-e\psi_0)/T}}.
 \label{Davies}
\end{equation}
Combining Eqs. (\ref{Poisson}) with the Boltzmann profiles leads to the 
Poisson-Boltzmann equation,
\begin{equation}
  \frac{\pd^2\psi}{\pd x^2} = \frac{8\pi e c_b}{\varepsilon}
           \sinh \frac{e\psi}{T},
 \label{PB}
\end{equation}
for the equilibrium double-layer potential \cite{VO,AndelmanES}.
By means of of the Poisson-Boltzmann equation the Davies isotherm 
can be expressed as
\begin{equation}
  \phi^+_0 = \frac {\phi^+_b} {\phi^+_b + \left[ b\phi^+_0 + \sqrt{
           (b\phi^+_0)^2+1} \right]^2
           \e^{-(\alpha+\beta\phi^+_0)/T}},
 \label{Davies2}
\end{equation}
where $b \equiv  a^+/(4\phi^+_b\lambda)$,
and $\lambda \equiv (8\pi c_b e^2/\varepsilon T)^{-1/2}$ is the
Debye-H\"{u}ckel screening length \cite{DH}.
Similar to Sec.~\ref{nonionic} one can calculate
the equilibrium equation of state,
\begin{equation}
  \Delta\gamma = \left[ T\ln(1-\phi^+_0) + 
               (\beta/2) (\phi^+_0)^2 - (2T/b) 
               (\sqrt{(b\phi^+_0)^2+1} - 1) \right] / (a^+)^2.
 \label{eqstate2}
\end{equation}
For weak fields the electrostatic correction to the equation of state 
is quadratic in the coverage, thus merely modifying the lateral 
interaction term, and for strong fields it becomes linear in the coverage.

{\em{\bf Kinetic equations}} are derived using the same scheme as before,
\[
   \frac{\pd\phi^\pm}{\pd t} = \frac{(a^\pm)^3 D^\pm}{T} \frac{\pd}{\pd x} \left[
   \phi^\pm \frac{\pd}{\pd x} \left( \frac{\delta\Delta\gamma}{\delta\phi^\pm}
   \right) \right],
\]
where $D^\pm$ are the diffusion coefficients of the two ions.
This variational scheme yields in the bulk solution the 
Smoluchowski diffusion equations,
\begin{equation}
  \frac{\pd\phi^\pm}{\pd t} = D^\pm \frac{\pd}{\pd x}
      \left( \frac{\pd\phi^\pm}{\pd x} \pm \frac{e}{T}
      \phi^\pm \frac{\pd\psi}{\pd x} \right),
 \label{diffusion2}
\end{equation}
at the layer adjacent to the interface
\begin{equation}
  \frac{\pd\phi^\pm_1}{\pd t} = \frac{D^\pm}{a^\pm} \left(
        \left.\frac{\pd\phi^\pm}{\pd x}\right|_{x=a^\pm} \pm 
        \frac{e}{T} \phi^\pm_1 \left.\frac{\pd\psi}{\pd x}
        \right|_{x=a^\pm} \right) - \frac{\pd\phi^\pm_0}{\pd t},
 \label{dp1dt2}
\end{equation}
and, finally, at the interface itself
\begin{equation}
  \frac{\pd\phi^+_0}{\pd t} = 
          \frac{D^+}{(a^+)^2} \phi^+_1
          \left[ \ln \frac{\phi^+_1(1-\phi^+_0)}{\phi^+_0} + 
          \frac{\alpha}{T} + \left( \frac{\beta\phi^+_0}{T} 
          - \frac{4\pi l}{a^+} \right) \phi^+_0 \right].
 \label{dp0dt2}
\end{equation}
We have made use of the electrostatic boundary condition 
(\ref{neutral}) in order to replace an electrostatic barrier
term, $e(\psi_0-\psi_1)/T$, with the approximate term 
$(4\pi l / a^+) \phi^+_0$, where $l \equiv e^2/\varepsilon T$ is 
the  Bjerrum length (about 7 \AA \ for water at room temperature).

We neglect electrodynamic effects, so the Poisson equation continues to hold.
The kinetic equations just derived, along with the Poisson equation and 
the necessary boundary and initial conditions, can be solved numerically 
(a similar set of equations is solved in Ref.~\cite{Radke}).

The relaxation in the bulk solution, accounted for by the Smoluchowski 
equations (\ref{diffusion2}), has the time scale
\[
  \tau_e = \lambda^2/D,
\]
where $D$ is an effective ambipolar diffusion coefficient.
This time scale is typically very short (microseconds),
\ie, the bulk relaxation is by orders of magnitude faster than in 
the non-ionic case.
The relaxation at the interface [Eq.~(\ref{dp0dt2})], by contrast,
is slowed down by the electrostatic repulsion, and has a time scale of
\[
  \tau_k = \tau_k^{(0)} \exp [e(\psi_{0}+\psi_{1})/T]
           \simeq \tau_k^{(0)} [ (a^+/2\lambda) (\phi^+_{0,\eq}/\phi^+_b)]^4 
                \exp [ -(4\pi l/a^+) \phi^+_{0,\eq}],
\]
where $\tau_k^{(0)}$ denotes the kinetic time scale in the absence
of electrostatics [Eq.~(\ref{asymkin})].
In salt-free surfactant solutions the surface potential reaches values
significantly larger than $T/e$, and hence, the interfacial relaxation
is by orders of magnitude slower than in the non-ionic case.

This analysis leads us to the conclusion that ionic surfactants in 
salt-free solutions undergo {\em kinetically limited adsorption}.
Indeed, dynamic surface tension curves of such solutions do not
exhibit the diffusive asymptotic time dependence of non-ionic surfactants,
depicted in Fig.~1.
The scheme of Sec.~\ref{nonionic}, focusing on the diffusive transport
inside the solution, is no longer valid.
Instead, the diffusive relaxation in the bulk solution is practically 
immediate and we should concentrate on the interfacial kinetics, 
Eq.~(\ref{dp0dt2}).
In this case the sub-surface volume fraction, $\phi^+_1$, obeys the Boltzmann
distribution, not the Davies adsorption isotherm (\ref{Davies}), 
and the electric potential is given by the Poisson-Boltzmann theory.
By these observations Eq.~(\ref{dp0dt2}) can be expressed as a function of
the surface coverage alone,
\begin{equation}
  \frac{\pd\phi^+_0}{\pd t} = \left(\frac{D^+\phi^+_b}{(a^+)^2}\right) \frac
      {\exp [(4\pi l/a^+)\phi^+_0]} {[b\phi^+_0+\sqrt{(b\phi^+_0)^2+1}]^2}
       \left\{ \ln \left[ \frac{\phi^+_b(1-\phi^+_0)}{\phi^+_0} \right] +
      \frac{\alpha}{T} + \frac{\beta\phi^+_0}{T} - 2\sinh^{-1}(b\phi^+_0) \right\},
 \label{dp0dt2_kin}
\end{equation}
thus reducing the problem to a single integration.

Not only does the scheme for solving the kinetic equations differ from the
non-ionic case, but also the way to calculate the dynamic surface tension 
has to change.
In kinetically limited adsorption the variation of the free energy with
respect to the surface coverage does not vanish, and, therefore, the equation of
state (\ref{eqstate2}) is strictly invalid out of equilibrium.
We derive the expression for the dynamic surface tension in the kinetically 
limited case from the general functional (\ref{Dg2}) by assuming
quasi-equilibrium inside the bulk solution (\ie, using Boltzmann profiles 
and the Poisson-Boltzmann equation).
This gives
\begin{eqnarray}
  \Delta\gamma[\phi^+_0(t)] &=& \{ T [ \phi^+_0\ln(\phi^+_0/\phi^+_b) 
        + (1-\phi^+_0)\ln(1-\phi^+_0) ] - \alpha\phi^+_0 - (\beta/2)(\phi_0^+)^2 
       \nonumber \\
      & & + 2T [ \phi^+_0\sinh^{-1}(b\phi^+_0) - (\sqrt{(b\phi^+_0)^2+1} - 1)/b ]
                \} / (a^\pm)^2.
 \label{Dgkinetic}
\end{eqnarray}

Assuming high surface potentials ($b\phi^+_0\gg 1$), the function defined in
Eq.~(\ref{Dgkinetic}) becomes non-convex for 
$\beta/T > 2(2+\sqrt{3}) \simeq 7.5$, as demonstrated in Fig.~4.
If that is indeed the case, our model predicts an unusual time dependence for
the dynamic surface tension, as observed in experiments (Fig.~3).
We thus infer that the shape of the experimental dynamic surface tension
curves is a consequence of a kinetically limited adsorption brought about
by strong electrostatic interactions.
Physically, the non-convexity results from a competition between 
short-range and long-range interactions. 
It implies a sort of a two-phase coexistence, which suggests the 
following scenario: 
As the surface coverage increases, the system reaches a local free-energy
minimum leading to a pause in the adsorption
(the intermediate plateau of the experimental curves). 
This metastable state lasts until domains of the denser, global-minimum
phase are nucleated, resulting in further increase in coverage and decrease 
in surface tension.
A complete, quantitative treatment of such a scenario cannot be presented
within our current formalism, since it inevitably leads to a 
monotonically-decreasing free energy as a function of time, and hence, 
cannot account for nucleation \cite{Langer}.

A value of $\beta > 7.5T$ is somewhat large for the lateral attraction 
between surfactant molecules.
Experimental estimation of this parameter for common non-ionic
surfactants yields a few $T$ \cite{Lin2}.
Throughout the above calculations we have assumed, to a sort of a
zeroth approximation, that no counterions are adsorbed at the interface.
It can be shown that the presence of a small amount of 
counterions at the interface introduces a correction to the free energy 
which is, to a first approximation, quadratic in the surfactant coverage, \ie,
leading to an effective increase in lateral attraction \cite{JPC}.
The addition to $\beta$ due to the counterions turns out to be 
$[2\pi l a^-/(a^+)^2]T$, which may amount to a few $T$.
This addition accounts for the larger value of $\beta$ required for 
non-convexity.

\section{Ionic Surfactants with Added Salt}
\label{ionicsalt}

Finally, we consider the effect of adding salt to an ionic surfactant
solution.
For simplicity, and in accord with practical conditions,
we assume that the salt ions are much more mobile than the 
surfactant and their concentration exceeds that of the surfactant.
In addition, we take the salt ions to be monovalent and surface-inactive.

Under these assumptions we can neglect the kinetics of the salt
ions and reduce their role to the formation of a thin electric double
layer near the interface, maintaining quasi-equilibrium with the 
adsorbed surface charge.
We take the double-layer potential to be in the linear,
Debye-H\"{u}ckel regime \cite{AndelmanES,DH},
\[
  \psi(x,t) = \frac{4\pi e\lambda}{\varepsilon a^2} \phi_0(t) \e^{-x/\lambda},
\]
with a modified definition of the Debye-H\"{u}ckel screening length,
$\lambda \equiv (8\pi c_s l)^{-1/2}$, $c_s\gg c_b$ being the 
salt concentration (the superscript '+' is omitted from the surfactant symbols
in this section).

Substituting this double-layer potential in Eqs.~(\ref{diffusion2}) and 
(\ref{dp1dt2}) we obtain the kinetic equations in the bulk and at the layer
adjacent to the interface,
\begin{eqnarray}
  \frac{\pd\phi}{\pd t} &=& D \frac{\pd}{\pd x} \left(
         \frac{\pd\phi}{\pd x} - \frac{\phi_0\e^{-x/\lambda}}{2a^2\lambda^2 c_s}
         \phi \right),
  \label{diffusion3} \\
  \frac{\pd\phi_1}{\pd t} &=& (D/a) \left( \left. \frac{\pd\phi}{\pd x} 
         \right|_{x=a} - \frac{\phi_0}{2a^2\lambda^2 c_s} \phi_1 \right) -
         \frac{\pd\phi_0}{\pd t}.
  \label{dp1dt3}
\end{eqnarray}
The kinetic equation at the interface itself  remains the same as (\ref{dp0dt2}).

Considering the electric potential as a small perturbation, Eqs.~(\ref{diffusion3})
and (\ref{dp1dt3}) lead to the asymptotic expression
\begin{eqnarray}
  \phi_1(t)/\phi_b &\simeq& 1 - \phi_{0,\eq}/(2a^2\lambda c_s)
           - \sqrt{\tau_d/t}; \ \ \ \ \ t\rightarrow\infty
    \nonumber \\
  \tau_d &\equiv& \tau_d^{(0)} \left[ 1 - \frac{c_b}{2c_s} - \frac{\phi_{0,\eq}}
          {2a^2\lambda c_s} \left( 1 - \frac{3c_b}{2c_s} \right) \right]^2,
 \label{asymdiffsalt}
\end{eqnarray}
where $\tau_d^{(0)}$ denotes the diffusion time scale in the non-ionic 
case [Eq.~(\ref{asymdiff})].
As expected, the screened electrostatic interactions introduce a small correction
to the diffusion time scale.
This correction decreases with increasing salt concentration.
            
Since the kinetic equation at the interface is identical to the one
in the absence of salt, so is the expression for the corresponding time scale.
However, in the case of added salt the electrostatic interactions are screened,
the surface potential is much smaller than $T/e$, and, therefore, the kinetic
time scale, $\tau_k$, is only slightly larger than the non-ionic one 
[Eq.~(\ref{asymkin})].

We infer that ionic surfactants with added salt behave much like
non-ionic surfactants, \ie, undergo diffusion-limited adsorption provided 
that no additional barriers to adsorption exist.
The departure from the non-ionic behavior depends on the salt concentration
and is described to first approximation by Eq.~(\ref{asymdiffsalt}).
The ``footprint'' of diffusion-limited adsorption, \ie, the $t^{-1/2}$
asymptotic time dependence, is observed in experiments, as
demonstrated in Fig.~5.
Consequently, the scheme described in Sec.~\ref{nonionic} for solving the 
adsorption problem and calculating the dynamic surface tension in the
non-ionic case, is applicable also to ionic surfactants in the presence of 
salt, and good fitting to experimental measurements can be obtained
\cite{Langevin}.
         
\section{Summary}
\label{summary}

We have reviewed an alternative theoretical approach to the 
fundamental problem of the adsorption kinetics of surfactants.
The formalism we present is more complete and general than previous 
ones as it yields the kinetics in the entire system, both in the
bulk solution and at the interface, relying on a single 
functional and reducing the number of externally inserted 
assumptions previously employed.

Common non-ionic surfactants, not hindered by any high barrier
to adsorption, are shown to undergo diffusion-limited 
adsorption, in agreement with experiments.
In the non-ionic case our general formalism coincides with previous
ones and helps clarify the validity of their assumptions.
Strong electrostatic interactions in salt-free ionic surfactant
solutions are found to have a dramatic effect.
The adsorption becomes kinetically limited, which may lead to an
unusual time dependence, as observed in dynamic surface tension 
measurements.
Such a scenario cannot be accounted for by previous models.
Addition of salt to ionic surfactant solutions leads to screening
of the electrostatic interactions, and the adsorption becomes
similar to the non-ionic one, \ie, diffusion-limited.
The departure from the non-ionic behavior as the salt concentration
is lowered has been described by a perturbative expansion.

A general method to calculate dynamic surface tension is obtained
from our formalism.
In the diffusion-limited case it coincides with previous results
which used the equilibrium equation of state,
but in the kinetically limited case it produces different
expressions leading to novel conclusions.

Our kinetic model is restricted to simple relaxation processes,
where the free energy monotonically decreases with time.
In order to provide a quantitative treatment of more complicated
situations, such as the ones described in Sec.~\ref{ionicnosalt}
for certain ionic surfactants, a more accurate theory is required.

Finally, the approach presented here may be easily extended to 
more complicated systems.
This flexibility has been demonstrated in Sec.~\ref{ionicnosalt}
by introducing electrostatic interactions.
Solutions above the critical micelle concentration and adsorption
accompanied by lateral diffusion \cite{latdiff} are just two
examples for other interesting extensions.

\newlength{\tmp}
\setlength{\tmp}{\parindent}
\setlength{\parindent}{0pt}
{\em Acknowledgments}
\setlength{\parindent}{\tmp}

We are indebted to D.~Langevin and A.~Bonfillon-Colin for introducing
us to the problem and for further cooperation.
We benefited from discussions with M.-W. Kim.
Support from the German-Israeli Foundation (G.I.F.)
under grant No.~I-0197 and the US-Israel Binational Foundation (B.S.F.)
under grant No.~94-00291 is gratefully acknowledged.



\section*{Figure Captions}

\begin{itemize}

 \item[{\bf Fig.~1}]
   Diffusion-limited adsorption exhibited by non-ionic surfactants.
   Four examples for dynamic surface tension measurements are shown:
   decyl alcohol at concentration $9.49\times 10^{-5}$M (open circles)
   adapted from Ref.~\cite{AddHutch};
   Triton X-100 at concentration $2.32\times 10^{-5}$M (squares)
   adapted from Ref.~\cite{Lin1};
   C$_{12}$EO$_8$ at concentration $6\times 10^{-5}$M (triangles)
   and C$_{10}$PY at concentration $4.35\times 10^{-4}$M (solid circles),
   both adapted from Ref.~\cite{Hua2}.
   The asymptotic $t^{-1/2}$ dependence shown by the solid fitting 
   lines is a ``footprint'' of diffusion-limited adsorption.

 \item[{\bf Fig.~2}]
  (a) Dependence between surface tension and surface coverage in
       diffusion-limited adsorption [Eq.~(\ref{eqstate})].
       The values taken for the parameters match the example in (b).
  (b) Typical dynamic surface tension curve in diffusion-limited
      adsorption (reproduced from Ref.~\cite{Lin2}).
      The solution contains $1.586\times 10^{-4}$M decanol.
      The solid line is a theoretical fit using the following parameters:
      $a=4.86$ \AA, $\alpha=11.6 T$, $\beta=3.90 T$ (all three fitted 
      from independent equilibrium measurements), and
      $D=6.75\times 10^{-6}$ cm$^2$/s.

 \item[{\bf Fig.~3}]
  Dynamic interfacial tension between SDS aqueous solutions and
  dodecane, adapted from Ref.~\cite{Langevin}:
  $3.5\times 10^{-4}$M SDS without salt (filled circles); 
  $4.86\times 10^{-5}$M SDS with 0.1M NaCl (open circles).

 \item[{\bf Fig.~4}]
  Dependence between surface tension and surface coverage 
  in kinetically limited adsorption [Eq.~(\ref{Dgkinetic})].
  The values taken for the parameters are: 
  $a^+=8.1$ \AA, $\phi^+_b=1.1\times 10^{-4}$, $\alpha=14T$ and 
  $\beta=9T$.
  Such a curve should lead to the qualitative time dependence found
  in the salt-free case (see Fig.~3).
  
 \item[{\bf Fig.~5}]
   Diffusion-limited adsorption exhibited by ionic surfactants
   with added salt:
   Dynamic interfacial tension between an aqueous solution of 
   $4.86\times 10^{-5}$M SDS with 0.1M NaCl and dodecane 
   (open circles and left ordinate), adapted from Ref.~\cite{Langevin};
   Dynamic surface tension of an aqueous solution of $2.0\times 10^{-4}$M 
   SDS with 0.5M NaCl (squares and left ordinate),
   adapted from Ref.~\cite{Fainerman};
   Surface coverage deduced from Second Harmonic Generation 
   measurements on a saturated aqueous solution of SDNS with 2\% NaCl 
   (filled circles and right ordinate), adapted from Ref.~\cite{SHG}.
   The asymptotic $t^{-1/2}$ dependence shown by the solid fitting 
   lines is a ``footprint'' of diffusion-limited adsorption.

\end{itemize}


\pagebreak

\begin{figure}[p] 
\epsfysize=16\baselineskip
\centerline{\hbox{ \epsffile{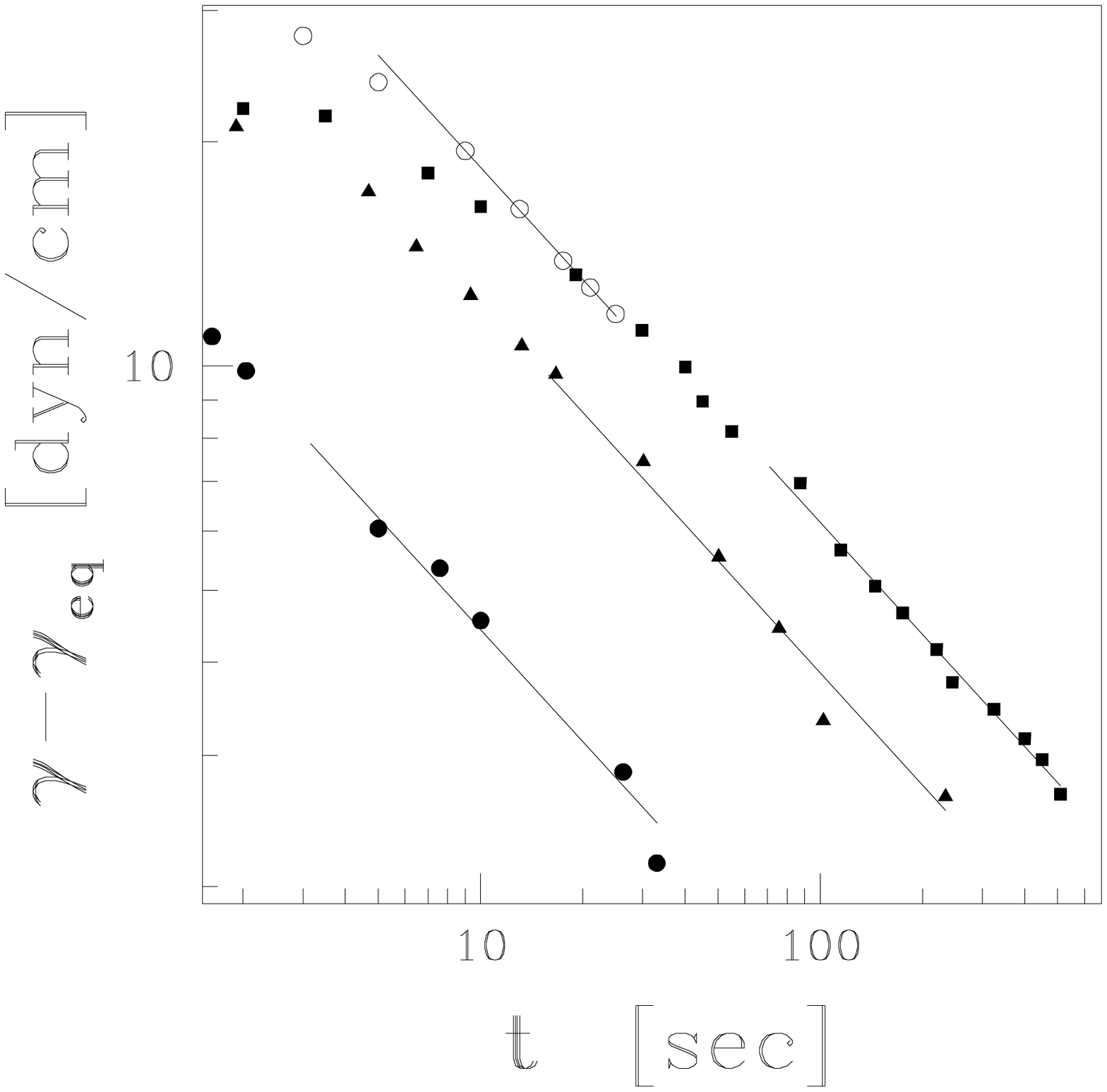} }} 
\caption[]{}
\end{figure}

\begin{figure}[p] 
\epsfysize=16\baselineskip
\centerline{\hbox{\epsffile{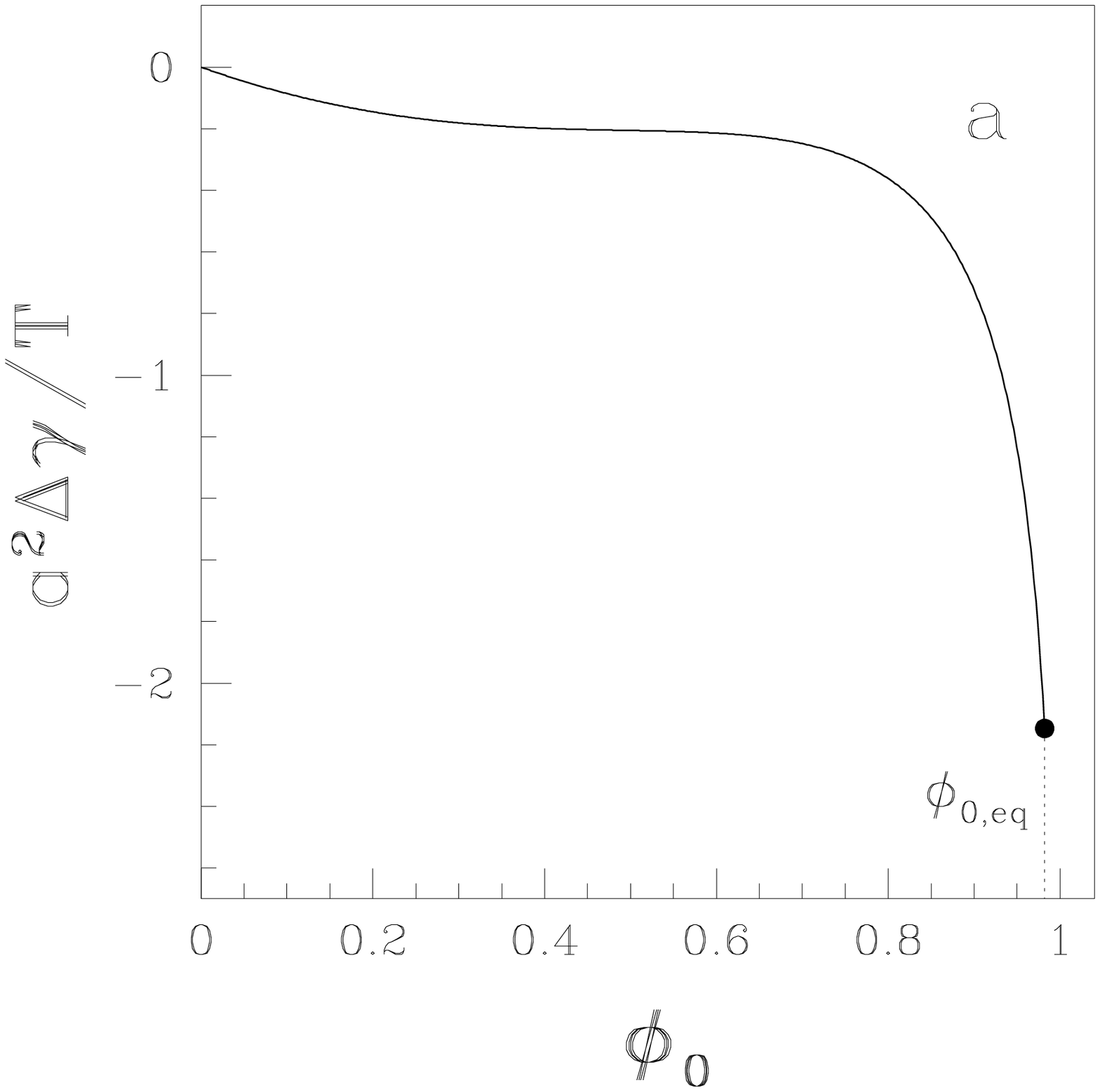}} 
                   \epsfysize=16\baselineskip \hbox{\epsffile{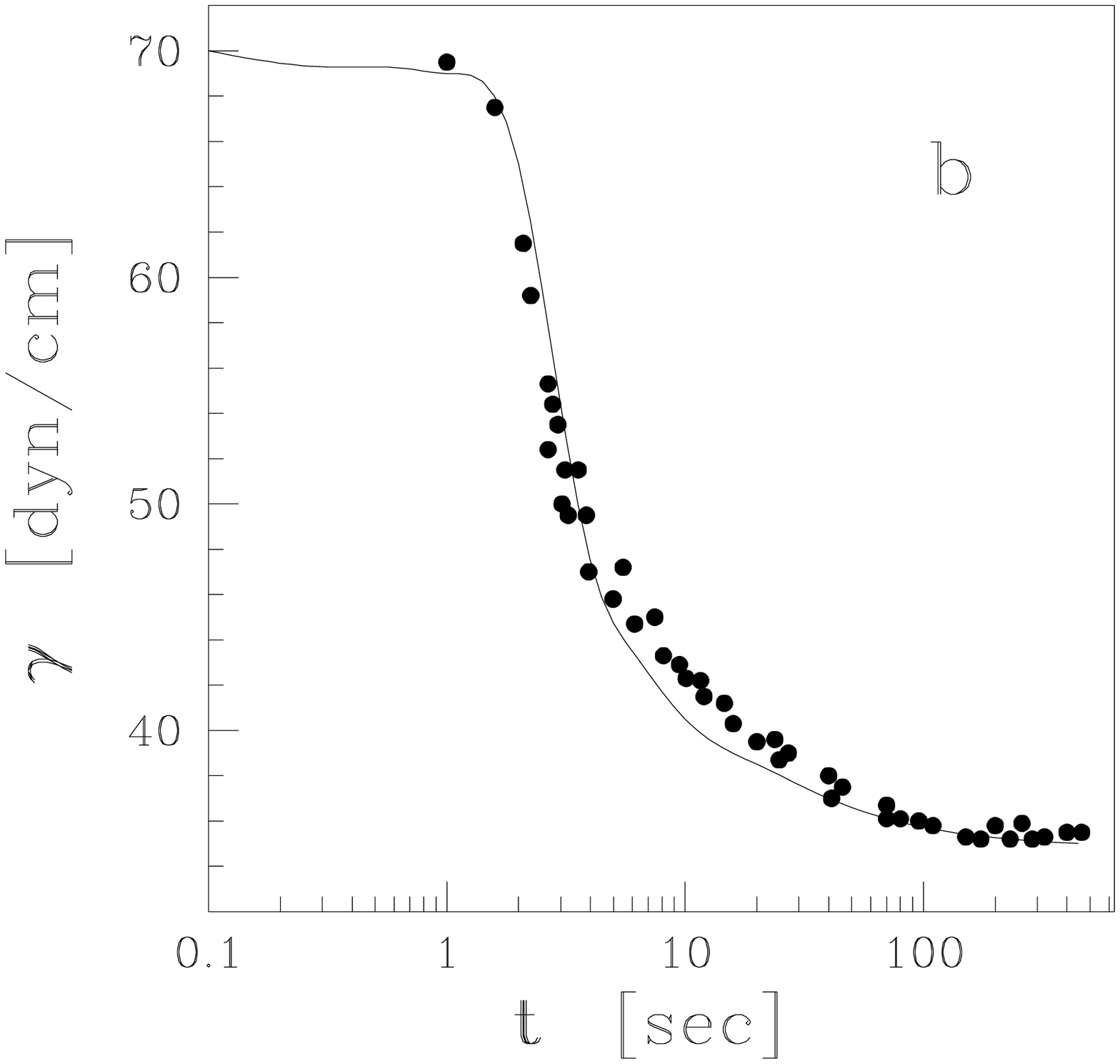}}}
\caption[]{}
\end{figure}  

\begin{figure}[p] 
\epsfysize=16\baselineskip
\centerline{\hbox{\epsffile{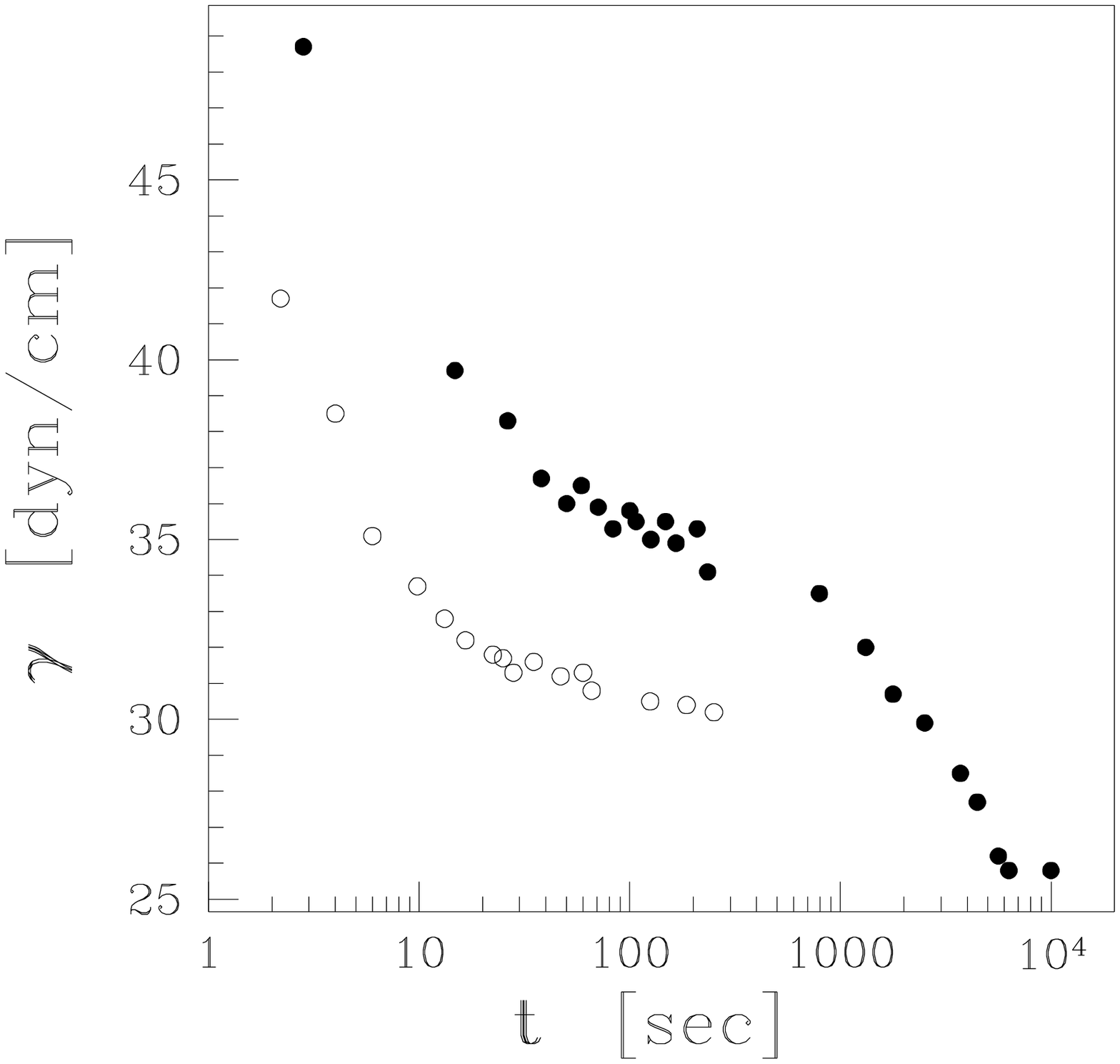}}}
\caption[]{}
\end{figure}

\begin{figure}[p] 
\epsfysize=16\baselineskip
\centerline{\hbox{\epsffile{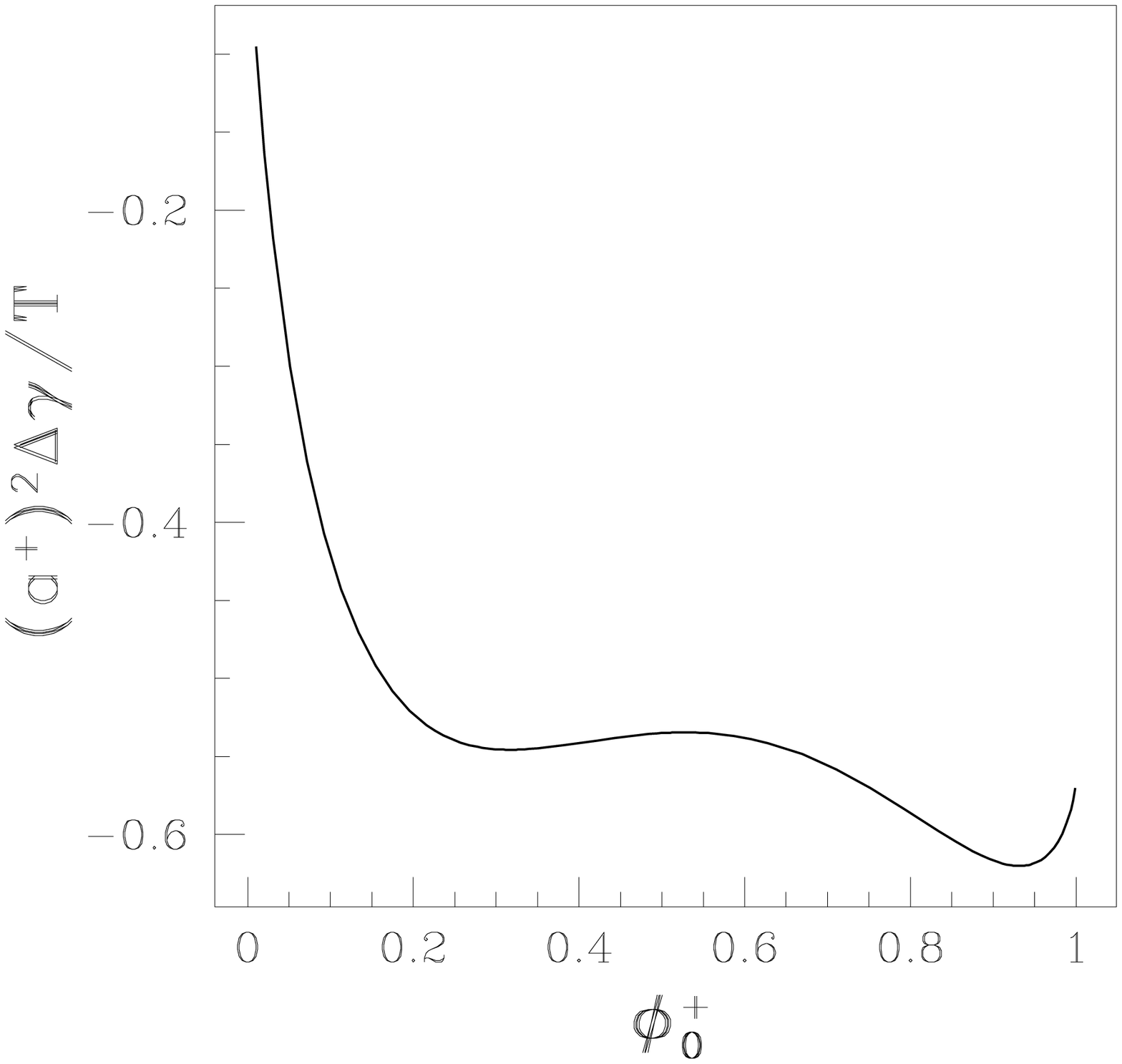}}}
\caption[]{}
\end{figure}

\begin{figure}[p] 
\epsfysize=16\baselineskip
\centerline{\hbox{ \epsffile{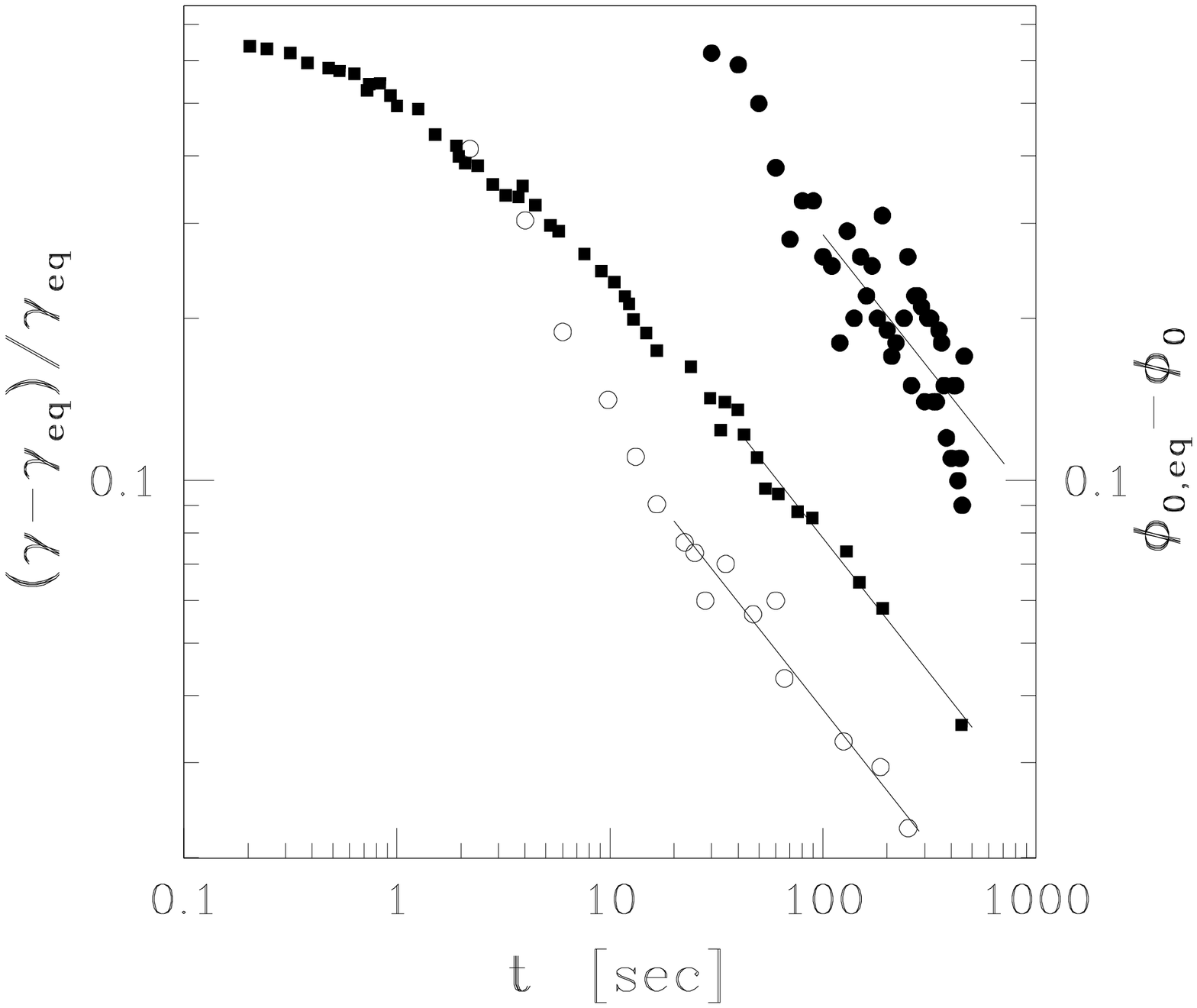} }} 
\caption[]{}
\label{diffcntlsalt}
\end{figure}



\begin{thebibliography}{99}

\bibitem{WT} Ward A F H, Tordai L (1946) 
{\it J Chem Phys} {\bf 14}: 453

\bibitem{review} For reviews of both experiments and theory see:
Borwankar R P, Wasan D T (1988) {\it Chem Eng Sci} {\bf 43}: 
1323;
Miller R, Kretzschmar G (1991) {\it Adv Colloid Interface Sci}
{\bf 37}: 97

\bibitem{Delahay} Delahay P, Fike C T (1958) 
{\it J Am Chem Soc} {\bf 80}: 2628

\bibitem{Hansen} Hansen R S (1960) {\it J Phys Chem}
{\bf 64}: 637

\bibitem{Joos1} van den Bogaert R, Joos P (1980)
{\it J Phys Chem} {\bf 84}: 190

\bibitem{Miller} Miller R, Kretzschmar G (1980)
{\it Colloid Polym Sci} {\bf 258}: 85

\bibitem{Borwasan1} Borwankar R P, Wasan D T (1983)
{\it Chem Eng Sci} {\bf 38}: 1637

\bibitem{Lin1} Lin S-Y, McKeigue K, Maldarelli C (1990)
{\it AIChE} {\bf 36}: 1785

\bibitem{Franses} Chang C H, Franses E I (1992)
{\it Colloids Surf} {\bf 69}: 189

\bibitem{Fordham} An earlier discussion of this assumption is
found in:
Fordham S (1954)
{\it Trans Faraday Soc} {\bf 54}: 593

\bibitem{EPL} Diamant H, Andelman D (1996)
{\it Europhys Lett} {\bf 34}: 575

\bibitem{JPC} Diamant H, Andelman D (1996)
{\it J Phys Chem} {\bf 100}: 13732

\bibitem{Langevin} Bonfillon A, Sicoli F, Langevin D (1993)
{\it Langmuir} {\bf 9}: 2172;
Bonfillon A, Sicoli F, Langevin D (1994) 
{\it J Colloid Interface Sci} {\bf 168}: 497

\bibitem{Tsonop} For an earlier discussion of such a distinction, see:
Tsonopoulos C, Newman J, Prausnitz J M (1971)
{\it Chem Eng Sci} {\bf 26}: 817

\bibitem{Adamson} Adamson A W (1990)
{\it Physical Chemistry of Surfaces}, 5th ed, Wiley \& Sons, 
New York, Chapters XI, XVI

\bibitem{Langer} See, for example,
Langer J S (1991) in: 
Godr\`{e}che C (ed) {\it Solids Far From Equilibrium},
Cambridge University Press

\bibitem{AddHutch} Addison C C, Hutchinson S K (1949) 
{\it J Chem Soc (London)}: 3387

\bibitem{Hua2} Hua X Y, Rosen M J (1991)
{\it J Colloid Interface Sci} {\bf 141}: 180

\bibitem{Sutherland} Sutherland K L (1952)
{\it Austral J Sci Res A} {\bf 5}: 683

\bibitem{Lin2} Lin S-Y, McKeigue K, Maldarelli C (1991)
{\it Langmuir} {\bf 7}: 1055

\bibitem{HuaDESS} Hua X Y, Rosen M J (1988)
{\it J Colloid Interface Sci} {\bf 124}: 652

\bibitem{Dukhin} Dukhin S S, Miller R, Kretzschmar G (1983) 
{\it Colloid Polym Sci} {\bf 261}: 335;
Miller R, Dukhin S S, Kretzschmar G (1985) 
{\it Colloid Polym Sci} {\bf 263}: 420

\bibitem{Borwasan2} Borwankar R P, Wasan D T (1986)
{\it Chem Eng Sci} {\bf 41}: 199

\bibitem{Radke} MacLeod C A, Radke C J (1994)
{\it Langmuir} {\bf 10}: 3555

\bibitem{Onsager} Onsager L, Samaras N N T (1934)
{\it J Chem Phys} {\bf 2}: 528

\bibitem{Davies} Davies J T (1958)
{\it Proc Roy Soc A} {\bf 245}: 417

\bibitem{VO} Verwey E J W, Overbeek J Th G (1948) 
{\it Theory of the Stability of Lyophobic Colloids}, 
Elsevier, New York

\bibitem{AndelmanES} Andelman D (1995) In: 
Lipowsky R, Sackmann E (eds) {\it Handbook of Biological Physics}. 
Elsevier, Amsterdam, Vol 1B

\bibitem{DH} Debye P, H\"{u}ckel E (1923)
{\it Phyzik} {\bf 24}: 185;
Debye P, H\"{u}ckel E (1924)
{\it Phyzik} {\bf 25}: 97

\bibitem{Fainerman} Fainerman V B (1978)
{\it Colloid J USSR} {\bf 40}: 769

\bibitem{SHG} Rasing Th, Stehlin T, Shen Y R, Kim M W, Valint Jr P 
(1988) {\it J Chem Phys} {\bf 89}: 3386

\bibitem{latdiff} There are cases encountered in practice where
lateral diffusion seems to play an important role. See:
Joos P, Fang J P, Serrien G (1992) 
{\it Colloid Interface Sci} {\bf 151}: 144;
Menger F M, Littau C A (1993)
{\it J Am Chem Soc} {\bf 115}: 10083

\end{thebibliography}
\end{document}